\pdfoutput=1

\documentclass[11pt]{article}

\usepackage[final]{acl}

\usepackage{times}
\usepackage{latexsym}

\usepackage[T1]{fontenc}

\usepackage[utf8]{inputenc}

\usepackage{microtype}
\usepackage{inconsolata}
\usepackage{amssymb}
\usepackage{dsfont}
\usepackage{amsmath}
\usepackage{booktabs}
\usepackage{multirow}
\usepackage[normalem]{ulem}
\useunder{\uline}{\ul}{}
\usepackage{amssymb}
\usepackage{bbm}
\usepackage{graphicx}
\usepackage{colortbl}
\usepackage{amsthm}
\usepackage{mathrsfs}
\usepackage{makecell}
\usepackage[linesnumbered,ruled,vlined]{algorithm2e}

\title{On the Role of Long-tail Knowledge in Retrieval Augmented Large Language Models}

\author{Dongyang Li$^{1,2}$
\thanks{\ \ D. Li, J. Yan and T. Zhang contributed equally to this work.}, 
Junbing Yan$^{1,2}$\footnotemark[1], Taolin Zhang$^{2}$\footnotemark[1], Chengyu Wang$^{2}$\footnotemark[2], \textbf{Xiaofeng He}$^{1,3}$\thanks{\ \ Co-corresponding authors.},\\ \bf Longtao Huang$^{2}$, Hui Xue$^{2}$, Jun Huang$^{2}$\\
$^1$ School of Computer Science and Technology, East China Normal University \\
$^2$ Alibaba Group, 
$^3$ NPPA Key Laboratory of Publishing Integration Development, ECNUP\\
  {\tt  dongyangli0612@gmail.com},
 {\tt  \{yanjunbing.yjb, zhangtaolin.ztl, chengyu.wcy,} \\  
 {\tt kaiyang.hlt, hui.xueh, huangjun.hj\}@alibaba-inc.com}, 
 {\tt  hexf@cs.ecnu.edu.cn}
 }

\begin{document}
\maketitle
\begin{abstract}
Retrieval augmented generation (RAG) exhibits outstanding performance in promoting the knowledge capabilities of large language models (LLMs) with retrieved documents related to user queries.
However, RAG only focuses on improving the response quality of LLMs via enhancing queries indiscriminately with retrieved information, paying little attention to what type of knowledge LLMs really need to answer original queries more accurately.
In this paper, we suggest that long-tail knowledge is crucial for RAG as LLMs have already remembered common world knowledge during large-scale pre-training.
Based on our observation, we propose a simple but effective long-tail knowledge detection method for LLMs.
Specifically, the novel Generative Expected Calibration Error (GECE) metric is derived to measure the ``long-tailness'' of knowledge based
on both statistics and semantics.
Hence, we retrieve relevant documents and infuse them into the model for patching knowledge loopholes only when the input query relates to long-tail knowledge.
Experiments show that, compared to existing RAG pipelines, our method achieves over 4x speedup in average inference time and consistent performance improvement in downstream tasks.
\end{abstract}

\section{Introduction}

Large language models (LLMs), equipped with retrieval augmented generation (RAG), perform well in various tasks \cite{DBLP:journals/jmlr/IzacardLLHPSDJRG23,DBLP:conf/emnlp/ChengHBZLW0WDZ23,DBLP:conf/emnlp/ShaoGSHDC23}.
RAG retrieves supplement knowledge by retrievers and enhances prompts for LLMs by retrieved documents, in order to generate more accurate contents \cite{DBLP:conf/icml/BorgeaudMHCRM0L22,DBLP:conf/emnlp/ChengHBZLW0WDZ23,DBLP:conf/emnlp/ShaoGSHDC23}.
However, previous RAG works concentrate on improving the task performance, without fine-grained process of knowledge \cite{DBLP:conf/emnlp/WangPXMLSDKLXAC23,DBLP:conf/acl/TrivediBKS23}.
In this case, redundant computation is performed on well-learned common knowledge, which does not require further enhancement. 
Therefore, more consideration should be given to long-tail knowledge that LLMs really need, which rarely occurs during pre-training \cite{DBLP:conf/icml/KandpalDRWR23}.
\footnote{Note that Long-tail knowledge is in low individual sample frequencies but high aggregated quantities, which implies a certain amount of significance~\cite{DBLP:journals/ipm/Jansen07b}.} 


\begin{figure}[tb]
\centering
\includegraphics[width=\columnwidth]{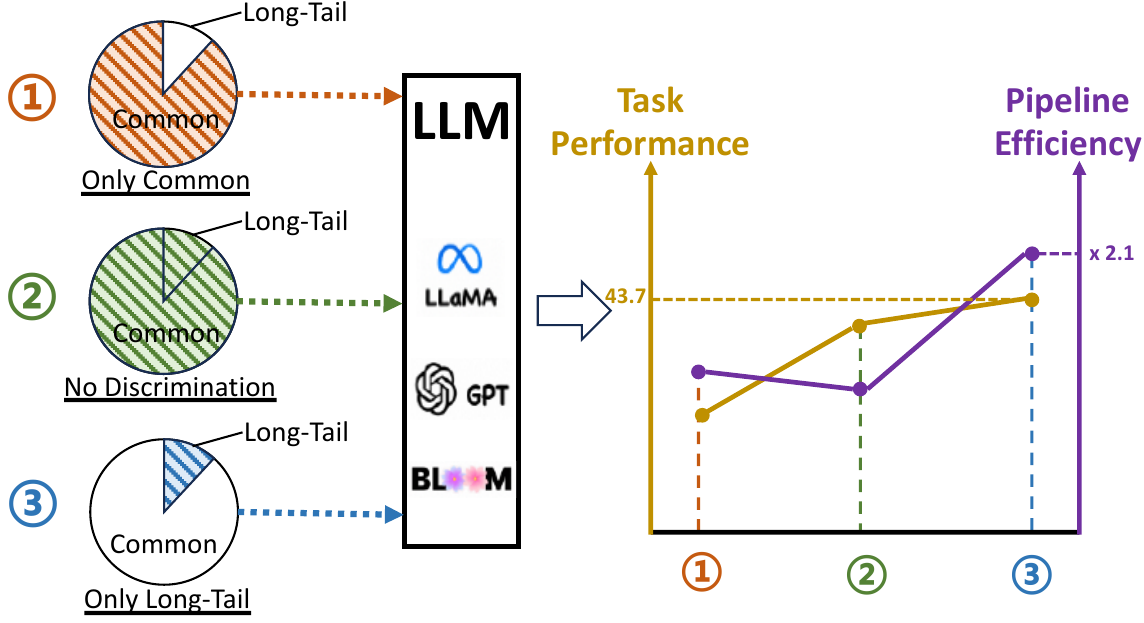} 
\caption{Comparison between different RAG strategies over the NQ dataset \cite{DBLP:journals/tacl/KwiatkowskiPRCP19}.}
\label{motivations}
\end{figure}

In the literature, RAG can be roughly divided into two categories: 
(1) \emph{Once Retrieval}. \citet{DBLP:conf/emnlp/WangPXMLSDKLXAC23,DBLP:conf/emnlp/ChengHBZLW0WDZ23,DBLP:journals/corr/abs-2301-12652} retrieve external knowledge just once by different retrievers and enhance the model with recalled related content for more effective generation. 
They treat all queries equally and do not model the familiarity of different queries to LLMs.
(2) \emph{Iterative Retrieval}. \citet{DBLP:conf/emnlp/ShaoGSHDC23,DBLP:journals/corr/abs-2310-05149,DBLP:journals/corr/abs-2310-11511} construct multi-step retrieval-then-augmentation process to generate accurate results by synergistic feedback of LLMs. 
Yet, as shown in Figure \ref{motivations}, augmenting LLMs with common knowledge that the models do not need results in low efficiency and redundant computation. 
To our knowledge, there is a lack of research on the use of long-tail knowledge for RAG.


Building upon our observation, we explore the role of long-tail knowledge in RAG. We suggest that long-tail knowledge is crucial for RAG and propose an improved RAG pipeline. 
Specifically, to measure the ``long-tailness'' of knowledge in terms of LLMs, we largely extend Expected Calibration Error (ECE) for classification tasks \cite{DBLP:conf/cvpr/AimarJFK23,DBLP:conf/cvpr/ZhongC0J21,DBLP:conf/nips/XuCY21}, and propose Generative Expected Calibration Error (GECE). It leverages METEOR \cite{DBLP:conf/acl/BanerjeeL05} and the output probability of LLMs to characterize ``long-tailness'', which considers both continuous gradient-based semantics and discrete frequency-based statistics.
Based on GECE, our pipeline retrieves relevant documents and performs RAG only when user queries relate to long-tail knowledge. 
Our approach outperforms current RAG pipelines, providing a 4x speedup in inference and improved performance in retrieval tasks.




\section{Related Work}
\subsection{Retrieval Augmentation}
The augmentation stage of RAG can be divided into three stages: pre-training, fine-tuning, and inference. Atlas \cite{DBLP:journals/jmlr/IzacardLLHPSDJRG23} is a retrieval-augmented pre-trained LLM and works well in few-shot settings. \citet{DBLP:conf/icml/BorgeaudMHCRM0L22,DBLP:conf/emnlp/WangPXMLSDKLXAC23} retrieve neighbor-related, chunk-grained knowledge from memory and inject the knowledge during the pre-training stage. 
\citet{DBLP:conf/emnlp/ChengHBZLW0WDZ23,DBLP:journals/corr/abs-2310-01352,DBLP:journals/corr/abs-2301-12652} fine-tune both the retriever and the generator synergistically and boost each other mutually.
\citet{DBLP:conf/emnlp/ShaoGSHDC23,DBLP:journals/corr/abs-2310-05149,DBLP:conf/acl/TrivediBKS23} insert knowledge at the inference stage by iterative guiding with frozen retrievers and LLMs. 
These methods introduce knowledge without detecting knowledge ``long-tailness'' and redundancy. 

\subsection{Long-Tail Processing}
\citet{DBLP:conf/icassp/ZhaoCCH23,DBLP:journals/tois/YaoZZS24,DBLP:conf/icassp/ZhengZHQ23} design repeat-sampling, under-sampling, and other strategies to access the unbalanced problem. 
They concentrate on classification tasks and consider less about the recent popular tendency of text generation tasks.
\citet{DBLP:journals/tkde/LiangLLZSG23,DBLP:journals/pr/ZhouZC23,DBLP:journals/tomccap/WangWYZYY24} leverage compositional operation to synthesize head and tail instances together by attention, graph-connection, and other fusion mechanisms. \citet{DBLP:conf/cvpr/WangZ0X23,DBLP:journals/pr/LiWHYZG23,DBLP:conf/aaai/XuXLLJY23} import extra features to tail classes for patching the demand of more information. 
To our knowledge, existing works touch less on distinguishing whether the instance is long-tail or not because of the existence of labeled training datasets. 

\section{Preliminaries}
\label{preliminary_label}

Traditional works rely on text frequencies to define whether the instance is long-tail or not; thus, low-frequency texts tend to be classified into long-tail classes.
For LLMs, computing text frequencies of previously unknown user queries is by no means an easy task.
As in \cite{DBLP:conf/cvpr/AimarJFK23,DBLP:conf/cvpr/ZhongC0J21,DBLP:conf/nips/XuCY21}, 
\emph{Expected Calibration Error} (ECE) provides a new perspective to measure ``long-tailness''.
ECE measures how well a model's estimated probabilities match true (observed) probabilities \cite{DBLP:conf/icml/GuoPSW17}. 
In the calculation of ECE, the confidence of each instance is allocated to a specific interval and obtained by the model predicted probability. 
The accuracy is determined by the comparison of the predicted label and the ground truth. 
The absolute margin between confidence and accuracy of each instance represents the calibration degree. 
The expected calibration degree of the whole dataset indicates the reliance of the model. 
Formally, ECE can be formulated as:
\begin{equation}
    ECE=\sum_{i=1}^{B} \frac{n_{b_{i}}}{N} \left | acc(b_i)-conf(b_i) \right | 
\end{equation}
where $i$ denotes $i$-th bin, $N$ is the total instance count of the dataset, $acc(b_i)$ and $conf(b_i)$ represent the accuracy and confidence of the bin $b_i$, and $n_{b_{i}}$ is the instance number of the bin $b_i$. $B$ is the count of bins in the interval of $[0, 1]$.
In our work, we extend ECE for NLP, particularly for the LLM text generation
scenario. 

\section{Methodology}

\subsection{Metric-based Long-tailness Detection}
As long-tail knowledge is crucial for RAG, we propose the GECE metric to detect the instance ``long-tailness''.
Here, we transform the traditional ECE formula with METEOR \cite{DBLP:conf/acl/BanerjeeL05} and average prediction probability:
\begin{itemize}
\item Accuracy in ECE is to measure the agreement between prediction and ground truth. In the generative scenario, we utilize METEOR \cite{DBLP:conf/acl/BanerjeeL05} to measure coherence and relevance between predicted candidates and ground truth.

\item Confidence in ECE is the predicted probability produced by the model itself. Similarly, we employ the average token probability output by LLMs.
\end{itemize}

Moreover, to enhance our metric with long-tail detection abilities, we further integrate the following two factors, which assist us to further separate common and long-tail instances apart:
\begin{itemize}
\item Average word frequency, as word frequency is a basic indication of long-tail texts.
\item Dot product between the mean gradient of the total dataset and the gradient of a specific instance is leveraged to evaluate the discrepancy~\cite{DBLP:journals/corr/abs-2201-05938}. This is because the gradient of a long-tail instance has a large disparity with the mean gradient of the total dataset, and vice versa.
\end{itemize}

From the above analysis, we construct GECE as:
\begin{equation}
    GECE=\frac{|M(pred, ref)-\frac{1}{n}\sum_{i=1}^{n}p(t_i) |}{\alpha\cdot[E(\bigtriangledown_{ins})\cdot \bigtriangledown_{ins} ] }  
\label{gece_equation}
\end{equation}
where $pred$ and $ref$ represent the generated text and the referenced ground truth, respectively. $M(pred, ref)$ is the METEOR score~\cite{DBLP:conf/acl/BanerjeeL05}.
The average token probability is formulated as $\frac{1}{n}\sum_{i=1}^{n}p(t_i)$ where $p(t_i)$ denotes the $i$-th token's probability produced by LLM, and $n$ is the token sequence length.
For the denominator part, $\alpha$ is the average word frequency. We can see that a long-tail instance has a smaller $\alpha$ value and hence its reciprocal will be larger. 
In addition, $\bigtriangledown_{ins}$ is the gradient w.r.t. the current instance, and $E(\bigtriangledown_{ins})$ is the mean gradient of the total dataset. To obtain the gradient, we run a forward and a backward pass only through fine-tuning the LLM using the dataset.
We can see that a long-tail instance has a smaller gradient $\bigtriangledown_{ins}$, compared to the mean score of the dataset, and thus obtains a smaller dot product $E(\bigtriangledown_{ins})\cdot \bigtriangledown_{ins}$.

Larger GECE value implies larger degree of long-tailness.
For example, if we apply GECE to the query of NQ
``Who was named African footballer of the year 2014'', the value is 34.6. In contrast,  for a long-tail, more professional NQ query ``Who has played Raoul in The Phantom of the Opera'', the GECE value is 112.7. 


\subsection{Improved RAG Pipeline}


As an extension to vanilla RAG pipelines, we only retrieve documents related to long-tail queries from the data source, disregarding common instances.
The retrieval process is implemented by a dense passage retriever to retrieve related WikiPedia\footnote{\url{https://www.wikipedia.org/}} documents.
For long-tail instances, we input the query concatenated with the recalled related documents to LLMs for answer attainment.
For common instances, we only input the query itself to LLMs.

\section{Experiments}

In this section, we briefly describe the experimental results and leave detailed experimental settings in Appendix \ref{experimental_settings}, and supplementary experimental results in Appendix \ref{Supplementary_Results}.

\begin{table*}[tb]
\centering
\small
\scriptsize
\resizebox{\linewidth}{!}{
\begin{tabular}{ccccccccccc}
\toprule
\multirow{2}{*}{\textbf{Model}} & \multirow{2}{*}{\textbf{Type}} & \multicolumn{4}{c}{\textbf{Rouge-1}}                    & \multicolumn{4}{c}{\textbf{Bleu-4}}                     & \multirow{2}{*}{\textbf{Speed-up}} \\
                                &                                & \textbf{10} & \textbf{15} & \textbf{20} & \textbf{Avg.} & \textbf{10} & \textbf{15} & \textbf{20} & \textbf{Avg.} &                                                                                               \\ \midrule
\multirow{2}{*}{Llama2-7B}      & w/o GECE                       & 41.2        & 42.2        & 42.9        & 42.1$_{(\pm 0.2)}$          & 7.19        & 7.31        & 7.40        & 7.30$_{(\pm 0.22)}$          & 1.0 $\times$                                                                                         \\
                                & w GECE                         & 41.9        & 43.1        & 43.7        & 42.9$_{(\pm 0.2)}$          & 7.27        & 7.40        & 7.48        & 7.38$_{(\pm 0.15)}$          &  2.1 $\times$                                                                                         \\ \midrule
\multirow{2}{*}{IRCoT}          & w/o GECE                       & 45.5        & 45.8        & 46.3        & 45.9$_{(\pm 0.3)}$          & 7.52        & 7.73        & 7.70        & 7.65$_{(\pm 0.31)}$          & 1.0 $\times$                                                                                         \\
                                & w GECE                         & 45.7        & 46.4        & 46.5        & 46.2$_{(\pm 0.3)}$          & 7.56        & 7.75        & 7.74        & 7.68$_{(\pm 0.26)}$          & 6.7 $\times$                                                                                         \\ \midrule
\multirow{2}{*}{SKR}            & w/o GECE                       & 46.3        & 47.0        & 47.2        & 46.8$_{(\pm 0.2)}$          & 7.57        & 7.65        & 7.79        & 7.67$_{(\pm 0.11)}$          & 1.0 $\times$                                                                                         \\
                                & w GECE                         & 46.9        & 47.1        & 47.6        & 47.2$_{(\pm 0.1)}$          & 7.66        & 7.78        & 7.85        & 7.76$_{(\pm 0.09)}$          & 5.5 $\times$                                                                                         \\ \midrule
\multirow{2}{*}{SELF-RAG}       & w/o GECE                       & 42.1        & 43.3        & 43.7        & 43.0$_{(\pm 0.3)}$          & 7.12        & 7.35        & 7.44        & 7.30$_{(\pm 0.28)}$          & 1.0 $\times$                                                                                         \\
                                & w GECE                         & 44.8        & 45.0        & 45.3        & 45.0$_{(\pm 0.2)}$          & 7.48        & 7.63        & 7.62        & 7.58$_{(\pm 0.22)}$          & 3.3 $\times$                                                                                         \\ \midrule
\multirow{2}{*}{FILCO}          & w/o GECE                       & 43.6        & 44.2        & 44.7        & 44.2$_{(\pm 0.3)}$          & 7.46        & 7.48        & 7.52        & 7.49$_{(\pm 0.17)}$          & 1.0 $\times$                                                                                         \\
                                & w GECE                         & 43.7        & 44.5        & 44.8        & 44.3$_{(\pm 0.2)}$          & 7.49        & 7.51        & 7.53        & 7.51$_{(\pm 0.15)}$          & 2.4 $\times$                                                                                         \\ \midrule
\multirow{2}{*}{ITER-RETGEN}    & w/o GECE                       & 45.5        & 46.4        & 47.1        & 46.3$_{(\pm 0.2)}$          & 7.63        & 7.75        & 7.78        & 7.72$_{(\pm 0.31)}$          & 1.0 $\times$                                                                                         \\
                                & w GECE                         & 46.5        & 47.0        & 47.3        & 46.9 $_{(\pm 0.1)}$         & 7.76        & 7.81        & 7.82        & 7.80$_{(\pm 0.26)}$          &  7.0 $\times$                                                                                         \\ \bottomrule
\end{tabular}
}
\caption{Experimental results on NQ. T-tests show the improvements are statistically significant with $p$ < 0.05.}
\label{general_results}
\end{table*}

\begin{table*}[tb]
\centering
\small
\scriptsize
\resizebox{\linewidth}{!}{
\begin{tabular}{ccccccccccc}
\toprule
\multirow{2}{*}{\textbf{Model}} & \multirow{2}{*}{\textbf{Type}} & \multicolumn{4}{c}{\textbf{Rouge-1}}                    & \multicolumn{4}{c}{\textbf{Bleu-4}}                     & \multirow{2}{*}{\textbf{Speed-up}} \\
                                &                                & \textbf{10} & \textbf{15} & \textbf{20} & \textbf{Avg.} & \textbf{10} & \textbf{15} & \textbf{20} & \textbf{Avg.} &                                    \\ \midrule
\multirow{2}{*}{Llama2-7B}      & w/o GECE                       & 22.5        & 24.6        & 24.9        & 24.0$_{(\pm 0.3)}$          & 6.68        & 6.92        & 7.17        & 6.92$_{(\pm 0.18)}$          & 1.0 $\times$                                \\
                                & w GECE                         & 23.3        & 25.2        & 25.8        & 24.8$_{(\pm 0.3)}$          & 6.74        & 6.99        & 7.25        & 6.99$_{(\pm 0.32)}$          & 2.2 $\times$                                \\ \midrule
\multirow{2}{*}{IRCoT}          & w/o GECE                       & 25.4        & 26.0        & 26.5        & 26.0$_{(\pm 0.2)}$          & 7.11        & 7.24        & 7.28        & 7.21$_{(\pm 0.24)}$          & 1.0 $\times$                                \\
                                & w GECE                         & 25.9        & 26.7        & 26.7        & 26.4$_{(\pm 0.1)}$          & 7.18        & 7.26        & 7.31        & 7.25$_{(\pm 0.17)}$          & 6.2 $\times$                                \\ \midrule
\multirow{2}{*}{SKR}            & w/o GECE                       & 26.6        & 27.2        & 27.5        & 27.1$_{(\pm 0.2)}$          & 7.51        & 7.57        & 7.62        & 7.57$_{(\pm 0.09)}$          & 1.0 $\times$                                \\
                                & w GECE                         & 27.1        & 27.3        & 27.6        & 27.3$_{(\pm 0.2)}$          & 7.54        & 7.60        & 7.63        & 7.59$_{(\pm 0.15)}$          & 6.0 $\times$                                \\ \midrule
\multirow{2}{*}{SELF-RAG}       & w/o GECE                       & 26.3        & 26.2        & 26.7        & 26.4$_{(\pm 0.2)}$          & 7.46        & 7.47        & 7.51        & 7.48$_{(\pm 0.19)}$          & 1.0 $\times$                                \\
                                & w GECE                         & 26.4        & 26.5        & 27.0        & 26.6$_{(\pm 0.1)}$          & 7.55        & 7.55        & 7.56        & 7.55$_{(\pm 0.26)}$          & 3.5 $\times$                                \\ \midrule
\multirow{2}{*}{FILCO}          & w/o GECE                       & 25.8        & 25.9        & 26.5        & 26.1$_{(\pm 0.3)}$          & 7.43        & 7.49        & 7.50        & 7.47$_{(\pm 0.16)}$          & 1.0 $\times$                                \\
                                & w GECE                         & 26.3        & 26.6        & 26.8        & 26.6$_{(\pm 0.1)}$          & 7.48        & 7.52        & 7.54        & 7.51$_{(\pm 0.23)}$          & 2.3 $\times$                                \\ \midrule
\multirow{2}{*}{ITER-RETGEN}    & w/o GECE                       & 26.8        & 26.7        & 27.2        & 26.9$_{(\pm 0.1)}$          & 7.36        & 7.41        & 7.57        & 7.45$_{(\pm 0.12)}$          & 1.0 $\times$                                \\
                                & w GECE                         & 27.1        & 27.3        & 27.4        & 27.3$_{(\pm 0.2)}$          & 7.49        & 7.55        & 7.59        & 7.54$_{(\pm 0.13)}$          & 7.3 $\times$                                \\ \bottomrule
\end{tabular}
}
\caption{Experimental results on TriviaQA. T-tests show the improvements are statistically significant with $p$ < 0.05.}
\label{general_results_TriviaQA}
\end{table*}

\subsection{Datasets}

\textbf{NQ} \cite{DBLP:journals/tacl/KwiatkowskiPRCP19} is a large-scale question answering dataset and constructed by human-labeled answers from Wikipedia web pages. We utilize the short answer type of NQ in this paper.
\textbf{TriviaQA} \cite{DBLP:conf/acl/JoshiCWZ17} is a relatively complex dataset containing syntactic and lexical differences between questions and answers.
\textbf{MMLU} \cite{DBLP:conf/iclr/HendrycksBBZMSS21} is a typical model evaluation benchmark that includes various-domain samples and it ranges in multiple degrees of difficulty from primary to advanced professional level.

\subsection{Baselines}
\textbf{Llama2-7B} \cite{DBLP:journals/corr/abs-2311-08377} is a pre-trained LLM with large-scale parameters and performs well on most benchmarks.
\textbf{IRCoT} \cite{DBLP:conf/acl/TrivediBKS23} introduces an interleaves retrieval approach, exploiting Chain-of-Thought (CoT) to assist the retrieval and leveraging the retrieval results to support CoT.
\textbf{SKR} \cite{DBLP:conf/emnlp/WangLSL23} utilizes LLMs to distinguish whether the query can be resolved or not, and only retrieve the knowledge out of the model's self-knowledge.
\textbf{SELF-RAG} \cite{DBLP:journals/corr/abs-2310-11511} introduces special reflection tokens to help the model to determine the retrieval requirement and retrieved content quality.
\textbf{FILCO} \cite{DBLP:journals/corr/abs-2311-08377} refines the retrieved context by a filter that is trained by string inclusion, lexical overlap relationship and conditional cross-mutual information.
\textbf{ITER-RETGEN} \cite{DBLP:conf/emnlp/ShaoGSHDC23} proposes a mutual promotion manner via the retrieval-augmented generation and generation-augmented retrieval.

\subsection{General Results}
We validate our method on the three datasets and the performance is listed in Table \ref{general_results}, Table \ref{general_results_TriviaQA}, and Table \ref{general_results_MMLU}. Due to space limitation, we move the result of MMLU to Appendix \ref{resutls_description_mmlu}.
From the results, we can observe that:
(1) All baseline models have better process speed when the data is filtered with GECE. Especially, the iterative methods are accelerated significantly (i.e., ITER-RETGEN and IRCoT). This improvement owes to the filter operation of GECE and the fine discrimination of the need or not for extra augmentation.
(2) With GECE, the task performance is also promoted by introducing less noise of the common instances.
(3) As the number of augmentation documents increases, i.e., from 10 to 20, the performance is boosted because of the substantial knowledge supplementation.

\subsection{Ablation Study}

\begin{table}[tb]
\centering
\small
\scriptsize
\resizebox{\linewidth}{!}{
\begin{tabular}{lccc}
\toprule
                   & \textbf{NQ}      & \textbf{TriviaQA} & \textbf{MMLU}     \\
                   & \textbf{Rouge-1} & \textbf{Rouge-1}  & \textbf{Accuracy} \\ \midrule
Ours               & 43.7             & 25.8              & 86.4              \\
Item Replacement  & 42.3             & 24.2              & 84.8              \\
\quad w/o Statistics only & 43.5             & 25.7              & 86.0              \\
\quad w/o Semantics only  & 41.6             & 24.9              & 85.5              \\ \bottomrule
\end{tabular}
}
\caption{Results of ablation study.}
\label{ablation_study}
\end{table}

In Table \ref{ablation_study}, (1) Item Replacement means that we utilize chrF \cite{DBLP:conf/wmt/Popovic15} and TER \cite{DBLP:conf/amta/SnoverDSMM06} to replace METEOR, two other metrics for text generation with the same value scale as METEOR. The replaced mean results of these two alternative metrics decline, indicating that METEOR is more accurate.
(2) For removing Statistics and Semantics, we delete the two items outside the absolute margin of GECE. The dropped scores demonstrate the importance of the two indicators.

\section{Conclusion}

In summary, our research highlights the significance of long-tail knowledge to enhance the efficacy of RAG for LLMs. We introduced the Generative Expected Calibration Error (GECE) to identify long-tail knowledge, which accelerates the inference process by more than fourfold in average and improves performance on downstream tasks without compromising the quality of responses. This demonstrates the benefits of selectively augmenting LLMs with targeted information, paving the way for more efficient and accurate RAG systems.

\section*{Acknowledgements}
We would like to thank anonymous reviewers for their valuable comments. This work was supported in part by National Key R\&D Program of China (No. 2022ZD0120302) and Alibaba Group through Alibaba Research Intern Program.

\section*{Limitations}
While our method shows considerable promise for improving the efficiency and accuracy of RAG-augmented language models, it is important to acknowledge several limitations. The long-tail knowledge detection method we propose is based on the GECE metric, which may not capture all dimensions of ``long-tailness''. Given that long-tail knowledge can be multi-faceted and context-specific, there may be instances where our method fails to detect, leading to suboptimal retrieval results. In addition, the applicability of GECE to more models and settings has not been thoroughly investigated. Further research is required to validate its effectiveness and adaptability across diverse LLMs and knowledge retrieval scenarios.

\section*{Ethical Considerations}
Our research on RAG for LLMs aims to enhance the precision and efficiency of knowledge retrieval, hence we believe that there are no direct negative social impacts associated with our contributions. Yet, it is important to acknowledge that any generative AI technology, including our application based on LLMs, must be deployed with careful consideration of its broader implications.



\bibliography{anthology,custom}
\bibliographystyle{acl_natbib}

\begin{table*}
\centering
\begin{tabular}{ccccccc}
\toprule
\multirow{2}{*}{\textbf{Model}} & \multirow{2}{*}{\textbf{Type}} & \multicolumn{4}{c}{\textbf{Accuracy}}                    & \multirow{2}{*}{\textbf{Speed-up}} \\
                                &                                & \textbf{10} & \textbf{15} & \textbf{20} & \textbf{Avg.} &                                    \\ \midrule
\multirow{2}{*}{Llama2-7B}      & w/o GECE                       & 84.9        & 85.4        & 85.5        & 85.3$_{(\pm 0.3)}$          & 1.0 $\times$                                \\
                                & w GECE                         & 85.3        & 86.1        & 86.4        & 85.9$_{(\pm 0.3)}$          & 2.4 $\times$                                \\ \midrule
\multirow{2}{*}{IRCoT}          & w/o GECE                       & 87.3        & 87.8        & 88.2        & 87.8$_{(\pm 0.5)}$          & 1.0 $\times$                                \\
                                & w GECE                         & 87.4        & 88.1        & 88.6        & 88.0$_{(\pm 0.4)}$          & 6.5 $\times$                                \\ \midrule
\multirow{2}{*}{SKR}            & w/o GECE                       & 87.8        & 89.2        & 89.6        & 88.9$_{(\pm 0.1)}$          & 1.0 $\times$                                \\
                                & w GECE                         & 89.2        & 89.6        & 89.7        & 89.5$_{(\pm 0.2)}$          & 6.3 $\times$                                \\ \midrule
\multirow{2}{*}{SELF-RAG}       & w/o GECE                       & 86.3        & 87.1        & 87.5        & 87.0$_{(\pm 0.4)}$          & 1.0 $\times$                                \\
                                & w GECE                         & 87.4        & 87.9        & 88.0        & 87.8$_{(\pm 0.3)}$          & 3.1 $\times$                                \\ \midrule
\multirow{2}{*}{FILCO}          & w/o GECE                       & 86.5        & 86.6        & 87.1        & 86.7$_{(\pm 0.2)}$          & 1.0 $\times$                                \\
                                & w GECE                         & 86.0        & 86.9        & 87.2        & 86.7$_{(\pm 0.3)}$          & 2.2 $\times$                                \\ \midrule
\multirow{2}{*}{ITER-RETGEN}    & w/o GECE                       & 88.7        & 89.5        & 89.4        & 89.2$_{(\pm 0.1)}$          & 1.0 $\times$                                \\
                                & w GECE                         & 89.2        & 89.6        & 89.8        & 89.5$_{(\pm 0.2)}$          & 7.1 $\times$                                \\ \bottomrule
\end{tabular}
\caption{Experimental results on MMLU. T-tests show the improvements are statistically significant with $p$ < 0.05. 
}
\label{general_results_MMLU}
\end{table*}

\newpage
\appendix

\section{Experimental Settings}
\label{experimental_settings}
For a fair comparison, we set baselines to the same backbone and retriever, i.e., Llama2-7B \cite{DBLP:journals/corr/abs-2311-08377} and DPR \cite{DBLP:conf/emnlp/KarpukhinOMLWEC20}, respectively. The utilization of GECE on SKR replaces the known/unknown judgment with GECE with other baseline operations set as usual. Our experiment results are averaged over multiple runs. The number of retrieved documents by DPR is set to \{10, 15, 20\}. The gradient of Equation \ref{gece_equation} is obtained from the average gradient of Feed-Forward Networks (FFN) in 29-32 layers. We categorize the instances with the top 20\% of large GECE values as long-tail instances and the rest as common instances.  The max related document token length is limited to 512. The temperature hyper-parameter of Llama2 is assigned as 0.6, top-p is set to 0.9.
Our ablation study is based on the baseline of Llama2-7B and the setting of 20 retrieved documents.

\section{Supplementary Experimental Results}
\label{Supplementary_Results}

\subsection{Additional Results on the MMLU Dataset} 
\label{resutls_description_mmlu}

The results over the MMLU dataset are shown in Table~\ref{general_results_MMLU}.
The conclusion is also consistent with the results over other datasets, showing the efficacy of the proposed method.

\subsection{Detailed Analysis of Statistics \& Semantics Information} 

\begin{figure}[tb]
\centering
\includegraphics[width=0.9\columnwidth]{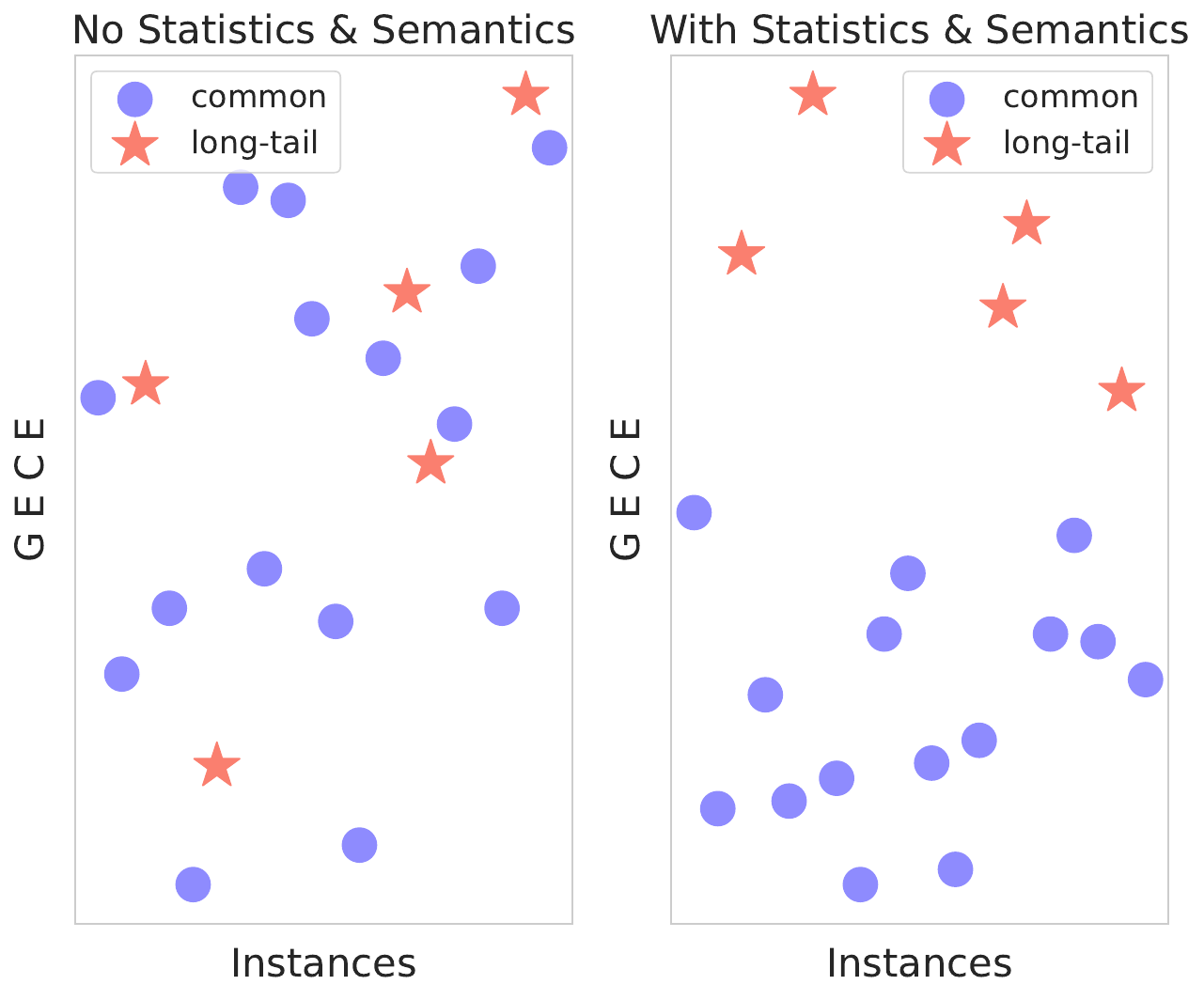} 
\caption{Comparison between absence and presence of statistics and semantics information in GECE.}
\label{statistics_and_semantics_comparison}
\end{figure}

To probe the influence of statistics and semantics information, we sample 15 common instances and 5 long-tail instances from  NQ and plot the GECE value of the sampled instance in Figure \ref{statistics_and_semantics_comparison}. Removing the statistics and semantics information leads to mixed and scattered instance distribution. With the help of the statistics and semantics information, we can separate common and long-tail instances apart distinctly.

\end{document}